\documentclass{article}

     \usepackage[preprint]{neurips_2020}

    \usepackage{amsfonts}
    \usepackage{amsmath}
    \usepackage{amssymb}

\usepackage[utf8]{inputenc} 
\usepackage[T1]{fontenc}    
\usepackage{hyperref}       
\usepackage{url}            
\usepackage{booktabs}       
\usepackage{amsfonts}       
\usepackage{nicefrac}       
\usepackage{microtype}      

\RequirePackage{framed,graphicx,multirow}
\usepackage{amssymb}
\usepackage{subfig}
\usepackage{bbm}
\RequirePackage{natbib}
\RequirePackage{amsthm}
\usepackage{multirow}
\usepackage{xcolor}
\usepackage{algorithm}
\usepackage[noend]{algpseudocode}

\title{Bayesian Estimation of the ETAS Model for Earthquake Occurrences.}

\graphicspath{{img/}{../img/}}

\author{%
  Gordon J.~Ross \\
  University of Edinburgh\\
  \texttt{gordon.ross@ed.ac.uk} \\
}

\begin{document}

\maketitle

\begin{abstract}
The Epidemic Type Aftershock Sequence  (ETAS) model is one of the most widely-used approaches to seismic forecasting. However most studies of ETAS use point estimates for the model parameters, which ignores the inherent uncertainty that arises from estimating these from historical earthquake catalogs, resulting in misleadingly optimistic forecasts. In contrast, Bayesian statistics allows parameter uncertainty to be explicitly represented, and fed into the forecast distribution. Despite its growing popularity in seismology, the application of Bayesian statistics to the ETAS model has been limited by the complex nature of the resulting posterior distribution which makes it infeasible to apply on catalogs containing more than a few hundred earthquakes. To combat this, we develop a new framework for estimating the ETAS model in a fully Bayesian manner, which can be efficiently scaled up to large catalogs containing thousands of earthquakes. We also provide easy-to-use software which implements our method.

\end{abstract}

\section*{Introduction}

The Epidemic Type Aftershock Sequence (ETAS) point process model is widely used to quantify the degree of seismic activity in a geographical region over a time period $[0,T]$, and to forecast the occurrence of future earthquakes \citep{ogata_space-time_1998,helmstetter_comparison_2006,wang_standard_2010}. The key intutition behind ETAS is that earthquakes tend to cluster in both time and space. Due to this clustering, the probability of an earthquake occurring at time $t$ and spatial location $(x,y)$  depends on the previous seismicity $H_t$, and is defined by the ETAS conditional intensity function:

\vspace{-5mm}

\begin{equation}
\lambda(t,x,y|H_t) = \mu f(x,y) + \sum_{t_i < t} Ke^{\alpha \left( m_i-M_0 \right)} h(t_i|c,p) s(x_i-x,y_i-y|\gamma), \quad  h(t_i|c,p) = \frac{(p-1)c^{p-1}}{(t-t_i+c)^p}.
\label{eqn:etas}
\end{equation}
where the summation is over all previous earthquakes that occurred in the region, with the $i^{th}$ such earthquake occurring at time $t_i$ and spatial location $(x_i,y_i)$, with magnitude $m_i$ drawm from the Gutenberg-Richter law with parameter $\beta$. The quantity $M_0$ denotes the magnitude of completeness of the catalog, so that $m_i \geq  M_0$ for all $i$. The parameter $\mu$ controls the background rate of seismicity, $K$ and $\alpha$ determine the productivity (average number of aftershocks) of an earthquake with magnitude $m_i$, and $c$ and $p$  are the parameters of the Modified Omori Law (which has here been normalized to integrate to 1) and represent the speed at which the aftershock rate decays over time. The function $f(x,y)$ is assumed to be known (typically a plug-in kernel density estimate is used \citep{helmstetter_comparison_2006}) and $s(x_i,y_i|\gamma)$ is a spatial density which determines how aftershocks are dispersed in space. Several choices have been studied in the literature, with the most common choices being the bivariate Gaussian and several types of power law distribution \citep{ogata_space-time_1998}:

\begin{equation}
s(x,y) = \frac{1}{2\pi \sigma_x \sigma_y} e^{-\frac{-x^2}{2\sigma_x^2} -\frac{y^2}{2\sigma_x^2}} \quad \mbox { or } \quad s(x,y) = \frac{(q-1)d^{q-1}}{\pi}\frac{1}{(x^2+y^2+d)^q}
\label{eqn:spatial}
\end{equation}

In cases where we are not interested in modelling the spatial distribution of earthquakes and are only concerned about their occurrence in time, we can set $s(x_i,y_i|\gamma) = 1$ and $f(x,y) = 1$ to obtain the temporal ETAS model. 

For  notational convenience we will write the unknown parameters as a vector $\theta = (\mu, K, \alpha, c, p, \gamma)$ (where $\gamma$ itself is a vector of multiple parameters defining the spatial kernel), and write $Y_i = (t_i,m_i,x_i,y_i)$ to denote the $i^{th}$ earthquake with $Y=(Y_1,\ldots,Y_n)$ denoting a particular catalog containing $n$ earthquakes. Most applications of the ETAS model to earthquake catalogs have been carried out within the frequentist statistical framework, where the unknown parameters $\theta$ are estimated using the maximum likelihood technique \citep{ogata_space-time_1998,veen_estimation_2008}, often making use of numerical approximations \citep{schoenberg_facilitated_2013,lippiello_parameter_2014}. This results in a single estimated value denoted by $\hat{\theta}$, which is then plugged into Equation \ref{eqn:etas} and treated as the true value for the purpose of forecasting future earthquakes. However when the ETAS model is estimated using a real earthquake catalog, the estimated value $\hat{\theta}$ will not be exactly equal to the true parameter $\theta$, and this misspecification can result in unreliable forecasts. Although this estimation error is most severe in shorter earthquake catalogs, it can also affect larger catalogs since as pointed out by \cite{veen_estimation_2008}, the flatness of the ETAS likelihood function makes several of the parameters difficult to accurately estimate.  While it is possible to compute the standard error of $\hat{\theta}$ under asymptotic assumptions \citep{wang_standard_2010}, it is difficult to translate this into information about forecast uncertainty without relying on computationally expensive methods such as bootstrapping \citep{fox_spatially_2016}.

Bayesian statistics represents an alternative statistical framework for reasoning about uncertainty, which is becoming increasingly popular in seismology \citep{omi_intermediate-term_2016,shcherbakov_bayesian_2014,holschneider_bayesian_2012},. In the Bayesian paradigm, we do not work with only  a single estimate of $\theta$ but instead consider the whole posterior distribution $p(\theta | Y)$ which represents our uncertainty about $\theta$ based on both the observed earthquake catalog and any prior knowledge we have based on previous studies. This uncertainty can then be incorporated into  forecasts in a straightforward manner \citep{glickman_basic_2003}. However despite its advantages, the Bayesian framework is difficult to apply since the posterior distribution in the ETAS model is highly complex. As such, even studies which attempt  Bayesian earthquake forecasting have had to resort to using frequentist-style point estimates for $\theta$, which mitigates the benefits of the Bayesian framework \citep{omi_intermediate-term_2016,ebrahimian_adaptive_2014}. The only attempts to at providing a fully Bayesian treatment of the ETAS model are \cite{vargas_bayesian_2012} and \cite{ebrahimian_robust_2017} which proposed using computational simulation based on the framework from \citet{rasmussen_bayesian_2013} for parameter estimation. However as we will show, their approach is likely to provide inaccurate parameter estimates and is not scalable to catalogs containing more than a few hundred earthquakes, which limits its applicability.

The purpose of this article is to introduce a new computational strategy for Bayesian estimation of the ETAS model based on a latent variable formulation which allows for efficient simulation from the posterior distribution even for large catalogs containing thousands of earthquakes. We have also implemented our estimation procedure in the \textbf{bayesianETAS} R package to allow practicing seismologists to automatically fit the Bayesian ETAS model without needing to understand the full details of the below mathematical calculations.


\section*{Direct Bayesian Estimation of the ETAS Model}
\label{sec:direct}

Suppose that the observations $Y=(Y_1,\ldots,Y_n)$ have been generated by a probability model $p(Y_1,\ldots,Y_n | \theta)$ with $\theta$ an unknown parameter vector. In Bayesian statistics, we begin with a prior distribution $\pi(\theta)$ which encodes all that is known about $\theta$ based on previous studies. In cases where we do not want previous knowledge to affect our analysis, $\pi(\theta)$ can be chosen to be non informative. After analyzing the data, the posterior distribution $p(\theta|Y_1,\ldots,Y_n)$ encodes all information about $\theta$ based on both the prior and the data, and is given by:

\begin{equation}
p(\theta|Y) = \frac{p(Y_1,\ldots,Y_n|\theta)\pi(\theta)}{\int p(Y_1,\ldots,Y_n|\theta)\pi(\theta)d\theta}.
\label{eqn:bayes}
\end{equation}

Knowledge of the posterior distribution allows point estimation of $\theta$ to be derived as in the maximum likelihood framework, but also allows all uncertainly about $\theta$ to be represented. This uncertainly can then be incorporated into forecasts by simply averaging the forecast distribution over the posterior. We do not intend to give a full treatment of Bayesian inherence here, but an interested reader can consult a standard reference such as \citet{glickman_basic_2003}.

Unfortunately in most real-world situations, the probability model will be too complicated to allow the integral in Equation \ref{eqn:bayes} to be computed analytically. As such, Bayesian inference typically involves using computer simulation to draw $M$ \textbf{samples} $\theta^{(1)},\ldots,\theta^{(M)} \sim p(\theta|Y)$ from the posterior distribution. These samples can then be used to compute all relevant quantities of interest. Although many sampling schemes exist, the most widely used is the independent random walk Markov Chain Monte Carlo (MCMC) method based on the Metropolis-Hastings algorithm. The simplest version of the MCMC method works as follows. First $\theta^{(1)}$ is initialized to a random value. Then, for each $2  \leq k \leq M$ a particular parameter $\theta^{(k)}_i$ is selected from the $\theta^{(k)}$ vector, and a new value $\theta^{(k+1)}_i$ is proposed as $\theta^{(k+1)}_i = \theta_i^{(k)} + \epsilon$ where $\epsilon \sim N(0,\sigma^2_i)$. This new value $\theta^{(k+1)}$ is accepted with probability $p(\theta^{(k+1)}|Y)/p(\theta^{(k)}|Y)$. If it is not accepted, then $\theta^{(k+1)}$ is replaced with the previous $\theta^{(k)}$ value. This procedure is then repeated until the desired number $M$ of $\theta^{(k)}$ values have been produced. It can be shown that these can be considered as a sample from the posterior distribution $p(\theta|Y)$ \citep{glickman_basic_2003}.



Based on the theory of point processes, the log-likelihood function for the ETAS model over a time period $[0,T]$ is \citep{veen_estimation_2008,ogata_space-time_1998}:

\vspace{-5mm}
\begin{equation}
\begin{split}
\log p(Y | \theta) = \sum_{i=1}^{n} \log & \left[\mu f(x,y)  + \sum_{j=1}^{i-1} \frac{(p-1)c^{p-1} K e^{\alpha(m_j-M_0)} g(x_i-x_j,y_i-y_j|\gamma)}{(t_i-t_j+c)^p} \right] - \\
&- \mu T - \sum_{i=1}^n Ke^{\alpha(m_i-M_0)} \left( 1- \frac{c^{p-1}}{(T-t_i+c)^{p-1}}\right).
\label{eqn:etaslikelihood}
\end{split}
\end{equation}
\vspace{-5mm}

Maximizing this function over $\theta$ gives the maximum likelihood estimate $\hat{\theta}$ used in frequentist inference. In the Bayesian framework, we are instead concerned with the posterior $p(\theta | Y ) \propto \pi(\theta)p(Y|\theta)$ where $\pi(\theta)$ is the prior which will be discussed later. The normalizing constant of this posterior cannot be computed analytically, and so simulation techniques must be used instead to draw samples from this distribution.. In \citet{vargas_bayesian_2012}, a simple independent random walk MCMC algorithm was introduced for this purpose. Given current values of the parameters $\theta^{(k)} =(\mu^{(k)}, K^{(k)}, \alpha^{(k)}, c^{(k)}, p^{(k)}, \gamma^{(k)})$, a new value  $\mu^{(k+1)} = \mu^{(k)} + \epsilon$ is proposed where $\epsilon \sim N(0,\sigma^2_\mu)$. This proposal is then accepted or rejected based on the standard Metropolis-Hastings ratio \citep{glickman_basic_2003}. The other parameters are then updated in a similar way, and the procedure is repeated until the desired number of samples has been drawn. We call this the direct approach to estimation, since it uses a standard MCMC algorithm based on the raw posterior.

Although this procedure is  simple and theoretically valid, there are grounds to doubt whether it will actually work well in practice. First, evaluating the likelihood function in Equation \ref{eqn:etaslikelihood} is an $O(n^2)$ operation due to the double summation and this evaluation must take place whenever a new parameter value is proposed in the MCMC algorithm. As such, it is computationally very demanding, and cannot feasibly be run on a catalog containing more than a few hundred earthquakes.  Second, in a seminal paper \cite{veen_estimation_2008} studied the performance of frequentist maximum likelihood estimation for the ETAS model based on directly maximizing the likelihood function in Equation \ref{eqn:etaslikelihood}, and found that the resulting parameter estimates often differed substantially from their true values. This is because the likelihood function is multi-modal and the components of $\theta$ are highly correlated. Since MCMC methods can also suffer from serious convergence issues when the parameters are correlated, it is reasonable to believe that this direct MCMC procedure will suffer from the same problem, and we demonstrate below that this is indeed the case. To avoid both these problems, we now introduce an alternative estimation scheme which allows for more reliable parameter estimation.

\section*{Latent Variable Formulation}
We now develop an alternative sampling posterior scheme based on introducing latent variables. These have the effect of breaking the dependence between the parameters in the likelihood function. We will show that conditional on the latent variables, the parameter sets $\{\mu\}$, $\{K,\alpha\}$, $\{p,c\}$ and $\{\gamma\}$ are all independent of each other, which greatly improves the convergence of MCMC sampling. 

 It has previously been shown \citep{rasmussen_bayesian_2013} that the ETAS model can be reinterpreted as a branching process in the following sense. Suppose that the $i^{th}$ earthquake occurs at time $t_i$, so that $i-1$ earthquakes have occurred previously. Equation \ref{eqn:etas} can be interpreted as showing that the ETAS intensity function at time $t_i$ is a sum of $i$ different Poisson processes. The first is a homogenous Poisson process with intensity $\mu f(x,y)$, while the other $i-1$ each correspond to one of the previous earthquakes. Specifically, for each $1 \leq j \leq i-1$, the earthquake at time $t_j$ triggers an inhomogeneous Poisson process with intensity $\lambda(t,x,y) = Ke^{\alpha(m_j-M_0) } (p-1)c^{p-1}(tt_j+c)^{-p}s(x-x_i,y-y_i)$. Based on standard results about the superposition of Poisson processes we can interpret event $t_i$ as having been generated by a single one of these $i$ processes. We hence introduce the latent branching variables $B=\{B_1,\ldots,B_n\}$ where $B_i \in \{0,1,\ldots,i-1\}$ indexes the process which generated $t_i$:

\vspace{-5mm}

$$
B_i \sim \left\{ \begin{array}{rl}
0 & \mbox{ if $t_i$ was produced by the background process (i.e. it is a mainshock) }\\
 j & \mbox{ if $t_i$ was triggered by the previous earthquake at time $t_j$ (i.e. it is an aftershock.)} \\
       \end{array} \right.
$$

\vspace{-5mm}

Conditional on knowing $B$, we can partition the earthquakes into $n+1$ sets $S_0,\ldots,S_n$ where $S_j = \{t_i ; B_i = j \}, \quad 0 \leq j  < n$ so that $S_0$ is the set of mainshock events which were not  triggered by previous earthquakes, and $S_j$ is the set of direct aftershocks triggered by the earthquake at time $t_j$. It is clear that these sets are mutually exclusive and that their union contains all the earthquakes in the catalog.  Additionally, we can see that the earthquakes in set $S_0$ are generated by a homogenous Poisson process with intensity $\mu f(x,y)$, while the events in each set $S_j$ for $j>0$ are generated by a single inhomogenous Poisson process with intensity $\lambda(t) = Ke^{\alpha(m_j-M_0) } (p-1)c^{p-1}(t-t_j+c)^{-p}s(x-x_i,y-y_i)$ . By multiplying together the likelihoods from each of these processes, the ETAS likelihood from Equation \ref{eqn:etaslikelihood} can hence be rewritten (conditional on knowing the branching variables) as:

\vspace{-5mm}

 \begin{align}
\begin{split}
p(Y | \theta, B)  = &   \left( \mu^{|S_0|} e^{-\mu T  }  \prod_{t_i \in S_0} f(x_i,y_i)\right) \times  \\
& \times \prod_{j=1}^{n} \left(e^{ -\kappa(m_j|K,\alpha)H(T-t_j|c,p)}   \kappa(m_j|K,\alpha)^{\left|S_j\right|}  \prod_{t_i \in S_j} h(t_i-t_j|c,p)s(x_i-x_j,y_i-y_j|\gamma)\right)  
\label{eqn:hawkesbayes}
 \end{split}
\end{align}
where $\kappa(\cdot)$ and $h(\cdot)$ are defined in Equation \ref{eqn:etas}, $|S_j|$ denotes the number of earthquakes in the set $|S_j|$, and $H(z) = \int_0^z h(t) dt = 1 - (c/(z+c))^{p-1}$. 
The key point of this reparameterization is that it makes both $\mu$ and $\gamma$ independent of the other model parameters in the posterior, while also drastically weakening the dependence between $(c,p)$ and $(K,\alpha)$. Indeed, this dependence is now restricted entirely to their interaction in the  $e^{ -\kappa(m_j|K,\alpha)H(T-t_j|c,p)}$ term. This greatly improves the performance of the MCMC sampler. In the next section we will discuss how this procedure  is carried out.

\subsection*{Parameter Estimation}
Our new MCMC scheme consists of sequentially sampling the parameters in the blocks $ \{\mu\}, \{K,\alpha\}, \{c,p\}, \{\gamma\}$  that are now only weakly dependent, given the latent variables $B$. Since the true values of  $B$ are unknown, they must also be estimated within the MCMC scheme. We begin by choosing arbitrary initial values $\theta^{)k)} = \left( \mu^{(1)},K^{(1)},\alpha^{(1)},c^{(1)},p^{(1)},B^{(1)}, \gamma^{(1)} \right)$ for the parameters. We then repeatedly sample new values $\theta^{(k+1)}$ from the posterior by repeatedly iterating the following four steps:

\begin{enumerate}

\item Sample a new value of $B^{(k+1)}$ from $p(B | Y, \theta^{(k)})$ from its exact conditional posterior. Assuming a uniform prior on each $B_i$, the probability of it being caused by any of the $i$ processes is simply the proportion of the overall intensity that can be attributed to that process, i.e.:
\begin{equation} 
p(B^{(k+1)}_i = j | Y, \theta) = \left\{ \begin{array}{rl}
\frac{\mu f(x_i,y_i)}{\mu f(x_i,y_i) + \sum_{j=1}^{i-1}   \kappa(m_j) h(t_i-t_j) s(x_i-x_j,y_i-y_j)} &\mbox{ if $j=0$} \\
\frac{\kappa(m_j) h(t_i-t_j)s(x_i-x_j,y_i-y_j) }{\mu f(x_i,y_i)+ \sum_{j=1}^{i-1}   \kappa(m_j) h(t_i-t_j) s(x_i-x_j,y_i-y_j)} &\mbox{ if $j \in {1,2,\ldots,i-1}$}
       \end{array} \right.
\end{equation}

Each $B_i$ can hence be drawn independently from the discrete distribution on $\{0,\ldots,i-1\}$, with weights given by the above. \vspace{2mm}


\vspace{2mm}

\item Sample a new value of $\mu^{(k+1)}$ from  $p(\mu | Y, \theta,B)$. Using Equation \ref{eqn:hawkesbayes} we can see this only depends on the events in the background process $S_0$:

\begin{equation}p \left( \mu | Y, \theta,B \right) \propto \pi(\mu)  e^{-\mu T} \mu^{|S_0|}
\end{equation}

This is equivalent to estimating the intensity function $\mu$ of a homogenous Poisson process on $[0,T]$, with event times $S_0$. In this case, the Gamma distribution is the conjugate prior: $\pi(\mu) = \mathrm{Gamma}(\alpha_\mu, \beta_\mu)$. The posterior distribution is then $p(\mu | Y, \theta,B) = \mathrm{Gamma} \left( \alpha_{\mu} + |S_0|, \beta_{\mu} + T \right)$ which can be sampled from directly.

\vspace{2mm}

\item Sample new values of $K^{(k+1)},\alpha^{(k+1)}$ from $p(K,\alpha | Y, \theta,B)$. Using Equation \ref{eqn:hawkesbayes}, we can see this is given by:

\begin{equation}p \left( K,\alpha | Y, \theta,B \right) \propto \pi(K,\alpha)  \prod_{j=1}^{n} e^{ -\kappa(m_j|K,\alpha)H(T-t_j|c,p)}   \kappa(m_j|K,\alpha)^{\left|S_j\right|}
\end{equation}

Although there is no conjugate prior in this case, it is straightforward to use random walk MCMC to draw a sample from this posterior.

\vspace{2mm}
\item Sample new values of $c^{(k+1)},p^{(k+1)}$ from $p ( c,p | Y, \theta,B )$. Using Equation \ref{eqn:hawkesbayes}, we can see this is given by:

\begin{equation}
p \left( c,p | Y, \theta,B \right) \propto  \pi(c,p) \prod_{j=1}^{n}  e^{ -\kappa(m_j|K,\alpha)H(T-t_j|c,p)} \prod_{t_i \in S_j} h(t_i-t_j|c,p)
\end{equation}
Again there is no conjugate prior but it is straightforwards to simulate from this distritbution using (e.g.) random walk MCMC.

\vspace{2mm}
\item Sample new values of $\gamma^{(k+1)}$ from:

\begin{equation}
p(\gamma | Y, \theta, B) \propto p(\gamma) \prod_{j=1}^n \prod_{t_i \in S_j} s(x_i-x_j,y_i-y_j | \gamma).
\end{equation}

For each earthquake $(t_i,m_i,x_i,y_i)$ which is not an immigrant (i.e. for which $B_i = j$ where $j \neq 0$) define $(x'_i,y'_i) = (x_i - x_{B_i}, y_i - y_{B_i})$ to be the recentered distance of earthquake $i$ from its triggering earthquake $j$. Assuming there are $n'$ non-immigrant earthquakes, the above expression becomes:

\begin{equation}
p(\gamma | Y, \theta, B) \propto p(\gamma) \prod_{j=1}^{n'} s(x'_i,y'_i | \gamma)
\end{equation}

In other words, the  $(x'_i,y'_i)$ locations are independent and identically distributed samples from $s(x,y)$, which allows for simple posterior inference.

\vspace{1mm}

In the case where $s(x,y)$ is the multivariate Normal distribution with a diagonal covariance matrix as in Equation \ref{eqn:spatial}, the parameters are $\gamma=(\sigma^2_x,\sigma^2_y)$ and the prior can be chosen to be a the conjugate Inverse-Gamma($\alpha_s,\beta_s$) distribution. In this case, the posteriors are simply $p(\sigma_x^2) = \mbox{Inverse-Gamma}(\alpha_s + n'/2, \beta_s + \sum_{i=1}^{n'} (x'_i)^2)$ and  $p(\sigma_y^2) = \mbox{Inverse-Gamma}(\alpha_s + n'/2, \beta_s + \sum_{i=1}^{n'} (y'_i)^2)$. When $s(x,y)$ is instead chosen to be a power law distribution as in \citep{ogata_space-time_1998} there will typically be no conugate prior, but random walk MCMC can be again be used to sample from this posterior.


\vspace{1mm}


\end{enumerate}

We note in passing that as well as providing estimates for the $(\mu, K, \alpha, c, p, \gamma)$ parameters of the ETAS model, this latent variable scheme also provides an estimate of the branching structure $\textbf{B}$ gives a declustering of the catalog into mainshocks and aftershocks, similar to the stochastic declustering introduced in \citep{zhuang_stochastic_2002}. However unlike the declustering introduced in the above reference, our Bayesian scheme provides a full posterior distribution over each $B_i$ and hence allows uncertainty about this branching structure to be quantified.

\section*{Performance Analysis}
\label{sec:synthetic}

To demonstrate the efficiency of our estimation scheme,we use it to estimate the ETAS model on the relocated Southern Californian earthquake catalog of \cite{hauksson_waveform_2012} (see the below Data snd Resources section). This catalog contains earthquakes from 1981 to 2019 in a rectangular region from $30^o$ to $37.5^o$ latitude and from $-113^o$ to $-122^o$ longitude. In order to compare the estimation efficiency on catalogs of difference sizes, we formed a number of subcatalogs containing only earthquakes above magnitude $M_0$. Specifically, we created catalogs of length $n \in (100,200,500,1000,2000,5000)$ by choosing $M_0 \in (5.02, 4.35, 4.04, 3.75, 3.36, 3.06)$. The larger catalog sizes here are more realistic than the $142$ earthquake catalog considered by  \cite{vargas_bayesian_2012} when testing their direct MCMC scheme.  Since our focus on this section is only on the computational efficiency of the MCMC schemes, we will keep our results here as general as possible and fit the temporal version of the ETAS model without a particular choice of the spatial kernel. However, we found essentially the same computational results for the spatial ETAS model using a variety of different kernels.

We use non informative priors for the ETAS parameters: $\mu$ is given a conjugate Gamma($0.1,0.1$) priors, and $K,\alpha,c$ are each assigned Uniform$(0,10)$. $p$ is given a Uniform$(1,10)$ prior to force it to be greater than $1$ \citep{holschneider_bayesian_2012}. Since these priors are all non-informative and are wide enough to cover all plausible parameter values that will be found on real earthquake catalogs, the estimation results are not sensitive to reasonable changes in these values.  For the proposal standard deviations used in the direct MCMC scheme, we chose values $\sigma_\mu=0.05, \sigma_K=0.15, \sigma_\alpha=0.15, \sigma_c = 0.25, \sigma_p = 0.30 $ based on a short pilot run of the simulation. These values result in an acceptance rate of between $20\%$ and $30\%$, which is thought to be optimal \citep{gelman_weak_1997}.


 Since the MCMC algorithm uses a random walk scheme, the resulting samples can be highly correlated which means they cannot be considered as independent draws from the posterior.  This means that the samples produced will be equivalent to a far smaller number of independent samples. As such, our main performance metric is the effective sample size (ESS) which measures how many independent samples the MCMC draws are equivalent to. Typically, a few hundred independent samples from the posterior are required for accurate inference.

\begin{table}[t]
\caption{Simulation Results}
\begin{center}

\begin{tabular}{c|ccccc|ccccc}
\hline
 & \multicolumn{5}{c|}{DIrect MCMC} & \multicolumn{5}{c}{Latent Variables} \\
n  & $\mu$ & K & $\alpha$ & c & p &  $\mu$ & K & $\alpha$ & c & p \\
 \hline
100 & 0.36 & 0.70 & 0.32& 0.25 & 0.32        &  0.03 & 0.16 & 0.03 & 0.02 & 0.06\\
200 & 1.86 & 3.59 & 0.87 & 1.58 & 10.26       &  0.18 & 1.3 & 0.09 & 0.19 & 0.69\\
500 & 7.49 & 10.23 & 6.92 & 13.49 & 20.62       &  0.97 & 2.24 & 0.34 & 2.43 & 6.10\\
1000 & 18.74 & 29.49 & 24.96 & 91.20 & 145.96        &  2.27 & 3.85 & 1.21 & 2.07 & 6.36\\
2000 & 337.01 & 96.56 & 96.02 & 290.80 & 342.06        &  2.38 & 2.72 & 2.93 & 2.32 & 2.95\\\
5000 & 2384.03 & 1143.01 & 1086.84 & 4594.58 & 1702.90        &  11.38 & 15.07 & 17.72 & 16.94 & 17.57\\
\hline
\end{tabular}
\caption*{Number of minutes required to draw samples equivalent to an effective sample size of $200$ when running on a catalog of length $n$. For example when the catalog contains 5,000 earthquakes, the direct scheme requires 2384 minutes (39 hours) to produce $200$ roughly independent samples of the $\mu$ parameter, compared to only $11.38$ minutes when using the latent variable approach}
\label{tab:results}
\end{center}
\end{table}

Table \ref{tab:results} shows the number of minutes that both the latent variable and direct schemes require to produce 200 effective samples for each catalog length, using the highly optimized C++ code from our \textbf{bayesianETAS} package, running on a Macbook Pro with an i7 2.4Ghz processor. It can be seen that as the catalog size increases, the direct method takes longer and longer to produce samples since carrying out a $O(n^2)$ likelihood evaluation per proposal, combined with the high degree of correlation in the parameters, seriously limits scalability. For the catalog containing 5,000 earthquakes, it takes around 4594 minutes (76 hours) to draw samples of all parameters equivalent to an effective sample size of only 200. In contrast, the latent variable scheme requires under 18 minutes to do the same. This is roughly around a 2500$\%$ improvement, and the size of the improvement grows with the length of the catalog. For catalogs containing more than 1,000 earthquakes, the direct scheme is hence not computationally feasible while the latent variable approach can provide a high number of posterior samples in a  reasonable length of time even for large catalogs.

\begin{figure*}[t]
\begin{center}
\captionsetup[subfigure]{labelformat=empty}

\subfloat[$\mu$]{
  \includegraphics[scale=0.12]{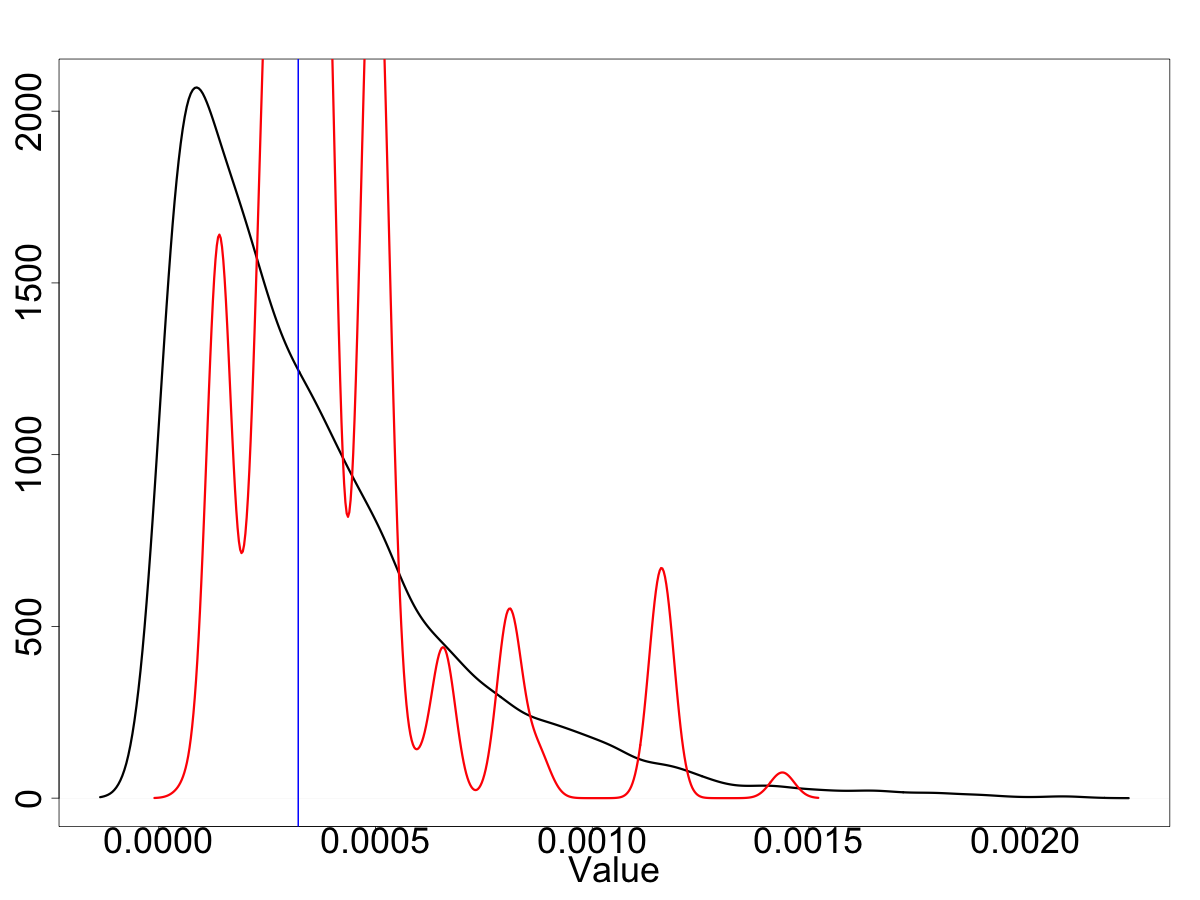}
}
\subfloat[K]{
  \includegraphics[scale=0.12]{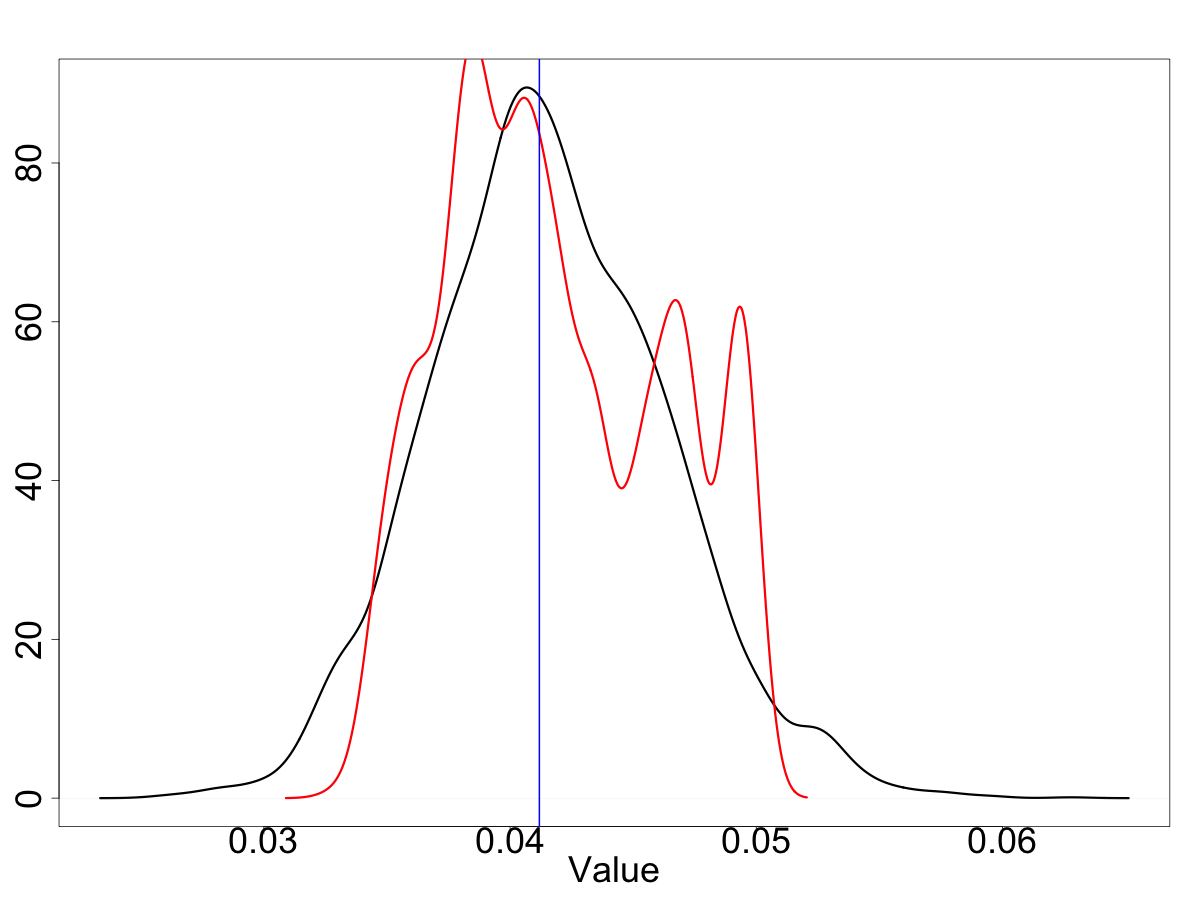}
}
\subfloat[$\alpha$]{
  \includegraphics[scale=0.12]{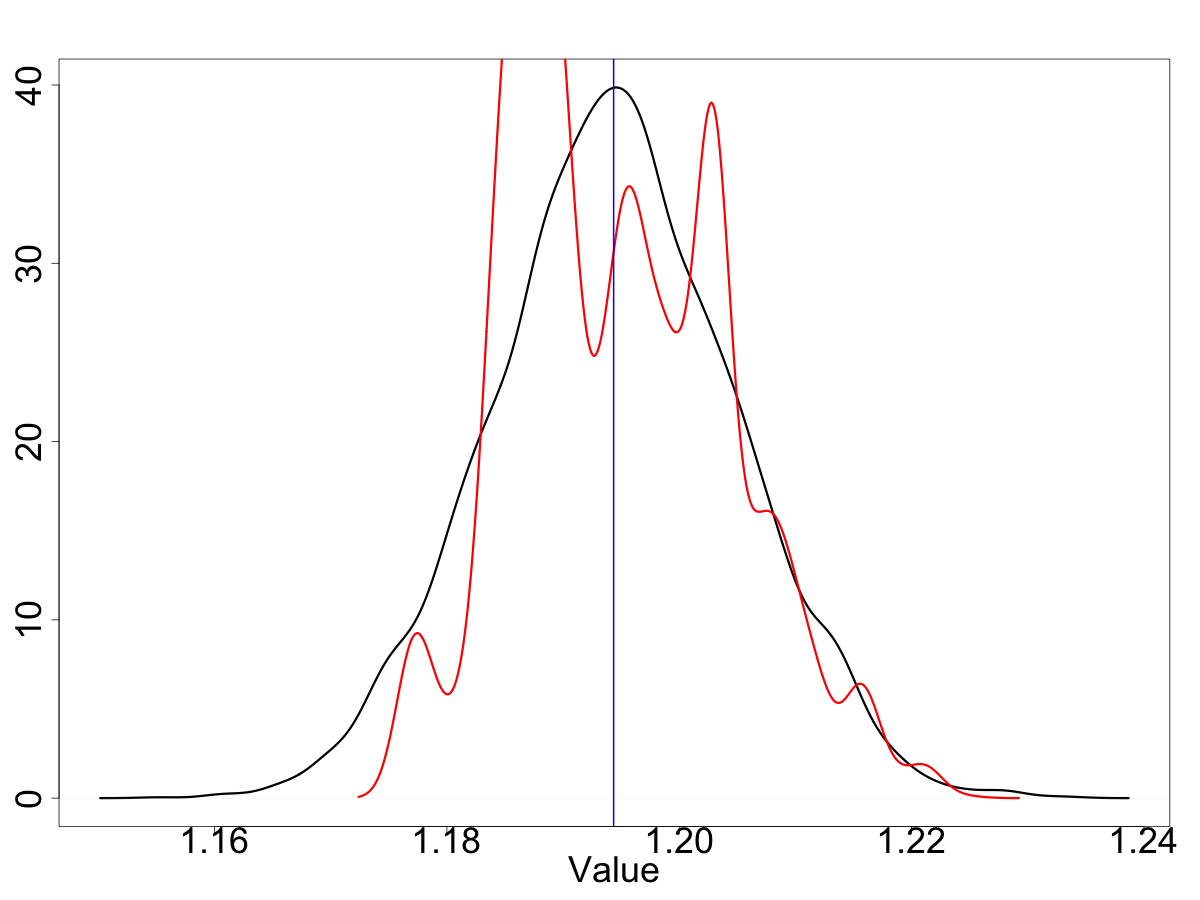}
}
\end{center}
\caption{Posterior distributions for three of the ETAS parameters on the simulated $1000$ event catalog using the direct MCMC method shown in red [grey in the printed version] and the latent variable approach shown in black. The maximum likelihood estimate of each parameter is shown as a verical line [blue in the online version] Similar patterns are observed for the other ETAS parameter posteriors but are omitted for space reasons.}
\label{fig:results}
\end{figure*}

To illustrate further, we consider a particular simulation run on the $5,000$ earthquake catalog taken from the above Southern Californian catalog.  For both the direct MCMC scheme and our latent variable formulation, 5,500 samples of each parameter were drawn from the posterior for each simulated catalog, with the first $500$ treated as a burn-in period and discarded. The direct method took $202$ minutes to complete, and resulted in an ESS of  $(17, 35, 37, 9, 24)$ for the parameters $(\mu,K,\alpha,c,p)$ respectively. while the latent variable method took $54$ minutes to produce an ESS of $(958, 723,  615, 643, 621)$. As such, we can see that the performance improvement is in both the overall running time, and the number of effective samples.  In fact, the low number of effective samples for the direct MCMC scheme is unlikely to allow for quantities such as forecast uncertainty to be computed with accuracy. To highlight this, Figure \ref{fig:results} plots a kernel density estimate of the resulting posterior distribution as computed by both methods.  It can be seen that the latent variable method produces a smooth posterior distribution which is expected given the much larger number of effective samples, while the direct approach suffers from high variability and multimodality due to the low effective sample size, and also underestimating the variance of $c$ and $p$. Even though the two sampling schemes are equivalent in the sense they would both converge to the same posterior given infinite computational run length, the latent variable scheme converges at a far faster rate which allows it to be deployed on catalogs containing thousands of events.

\section*{Forecasting}
\label{sec:forecasting}

A key advantage of using Bayesian inference to estimate the ETAS model is that it allows all uncertainty about the ETAS parameters to be incorporated when forecasting future earthquakes. This prevents the forecasts from being incorrectly overconfident due to ignoring this uncertainty. This is difficult to achieve when using maximum likelihood estimation even when standard errors for the parameters are available \citep{wang_standard_2010} since translating these into forecast uncertainty typically requires computationally expensive bootstrap procedures \citep{fox_spatially_2016}.

To forecast from the Bayesian ETAS model, suppose that we have observed earthquakes $Y$ on some time-interval $[0,T]$ and that we wish to forecast the occurrence of earthquakes on a future interval $[T,T+\delta]$. Denote these future earthquakes by $\tilde{Y}$. The general Bayesian forecast distribution is:

\begin{equation}
p(\tilde{Y} | Y) = \int p(\tilde{Y} | \theta)p(\theta|Y) d\theta \approx \sum_{k=1}^M p(\tilde{Y}|\theta^{(k)})
\end{equation}
where the $\theta^{(k)}$'s are the ETAS parameters that have been sampled from the posterior $p(\theta|Y)$ using the MCMC routine in the previous section. Particular forecasts can be easily produced by using simulation to approximate the relevant functions of this distribution. For example, if we wish to make a forecast for the average number of earthquakes that will occur in  $[T,T+\delta]$  on some spatial region $S$ then we can simulate $M$ realisations of the ETAS model on $[T,T+\delta]$ (one for each value of $\theta^{(k)}$ conditional on $Y$ and count the number of events in each. The resulting distribution is the forecast distribution for the number of earthquakes. Similarly, if we wish to compute the probability that an earthquake of magnitude greater than some threshold $R$ will occur during this time, then we can generate this as the proportion of simulated realisations which contain an earthquake with magnitude greater than $R$. Since these forecasts are using all of the information contained in the different values of $\theta^{(k)}$, they incorporates the parameter uncertainty which is lost using maximum likelihood based forecasting which relies only on a single estimated value of $\theta$.

\begin{figure}[t]
\begin{center}
  \includegraphics[scale=0.40]{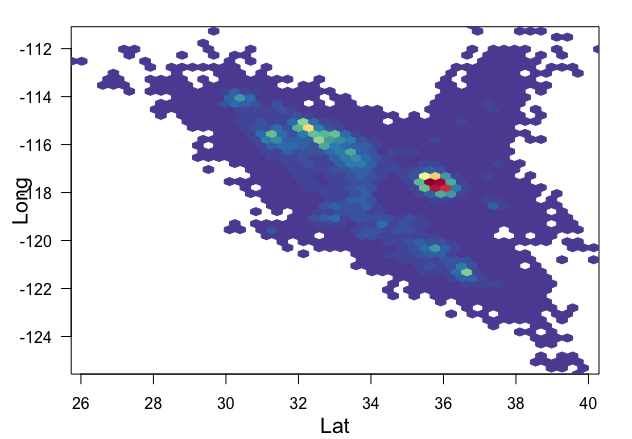}
  \end{center}
  \caption{Histogram plot of the forecasted density of earthquakes with magnitude greater than $5$ in Southern California, with dark blue [dark colouring in the printed version] corresponding to low density regions and red to high density. The strong red peak [lighter coloring in the printed version] in the centre-right of the plot is due to the large magnitude 7.1 earthquake which occurred at location $[-35.77, -117.60]$ in July 2019, which  still continues to produce aftershocks during the forecast period.}
  \label{fig:hazard}
\end{figure}

To illustrate this procedure, we fit the ETAS model to the entire 2010-2019 portion of the Southern California data-set described in the previous section using a bivariate Gaussian kernel for $s(x,y)$ as in Equation \ref{eqn:spatial}, with a non-informative Inverse-Gamma($0.1,0.1$) prior on both $\sigma^2_x$ and $\sigma^2_y$. A kernel density estimate for $f(x,y)$ was derived from the data prior to fitting the Bayesian model and held constant throughout, as discussed in \citep{helmstetter_comparison_2006}.  Figure \ref{fig:hazard} shows the resulting Bayesian forecast distribution of the likely locations for earthquakes with magnitude of greater than $5$ during the next one year after the end of the catalog. The strong red peak in the centre-right of the plot is due to the large magnitude 7.1 earthquake which occurred at location $[-35.77, -117.60]$ in July 2019, which  still continues to produce aftershocks during the forecast period.

\subsection*{Forecast Validation}

The previous section showed that the Bayesian ETAS model can produce spatial earthquake forecasts which take into account parameter uncertainty. In this section, we study the accuracy of these forecasts by attempting to predict the true number of earthquakes that occurred, in a retrospective analysis.

\begin{figure*}[t]
\begin{center}
\captionsetup[subfigure]{labelformat=empty}

\subfloat[$\mu$]{
  \includegraphics[scale=0.15]{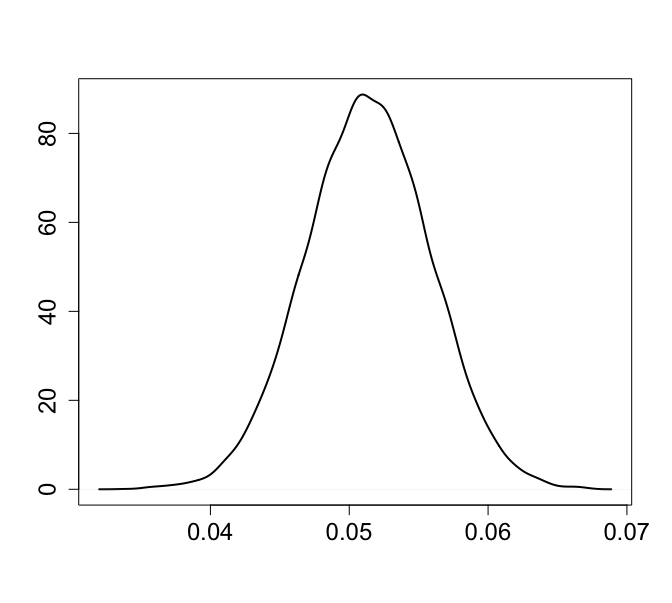}
}
\subfloat[K]{
  \includegraphics[scale=0.15]{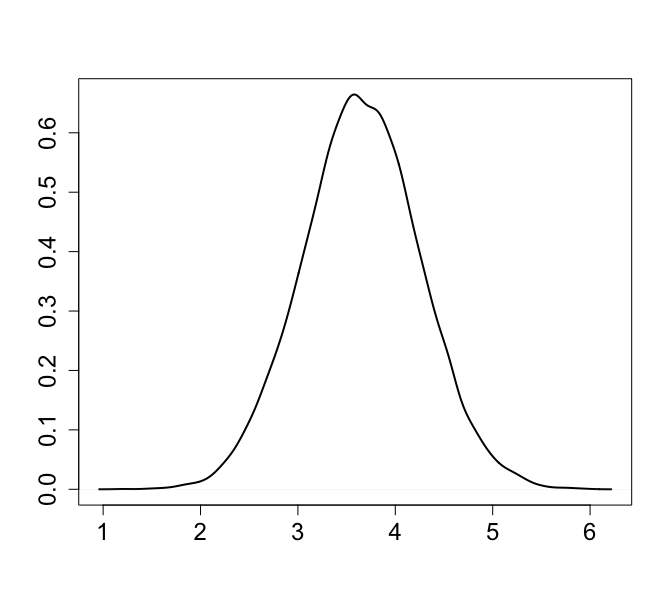}
}
\subfloat[$\alpha$]{
  \includegraphics[scale=0.15]{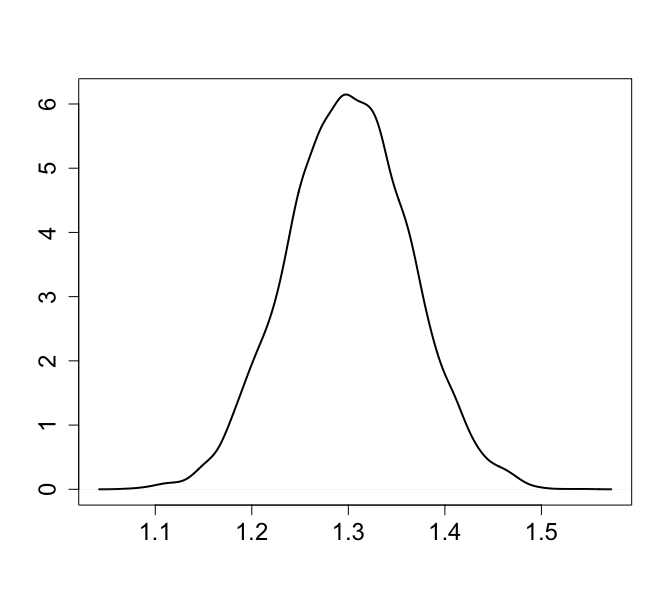}
}
\subfloat[c]{
  \includegraphics[scale=0.15]{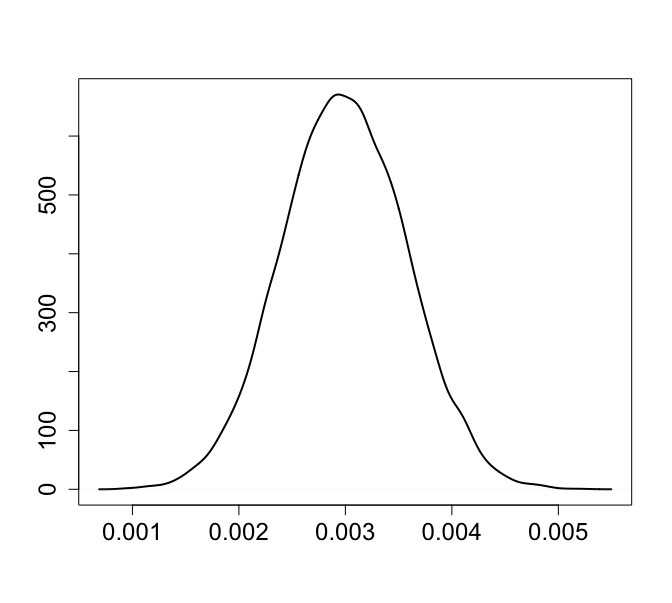}
} \\
\subfloat[p]{
  \includegraphics[scale=0.15]{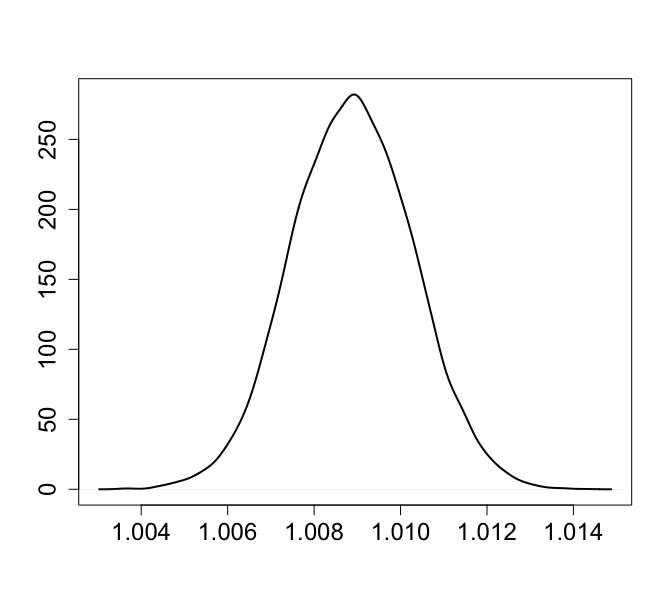}
}
\subfloat[$\sigma_x^2$]{
  \includegraphics[scale=0.15]{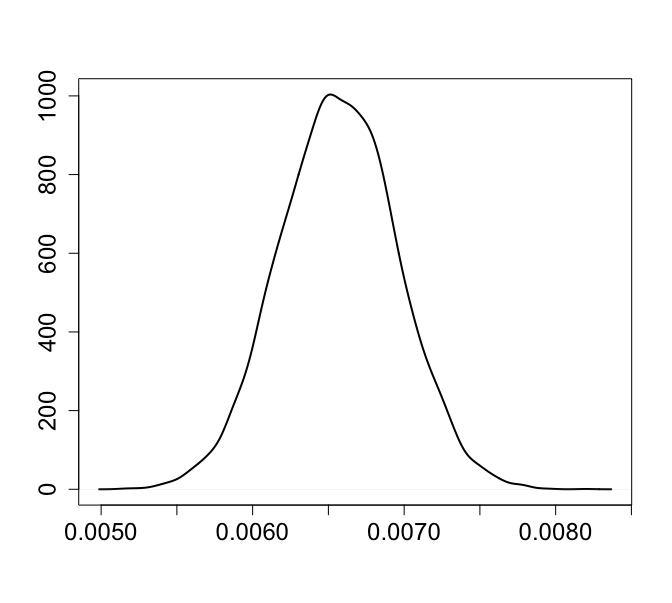}
}
\subfloat[$\sigma_y^2$]{
  \includegraphics[scale=0.15]{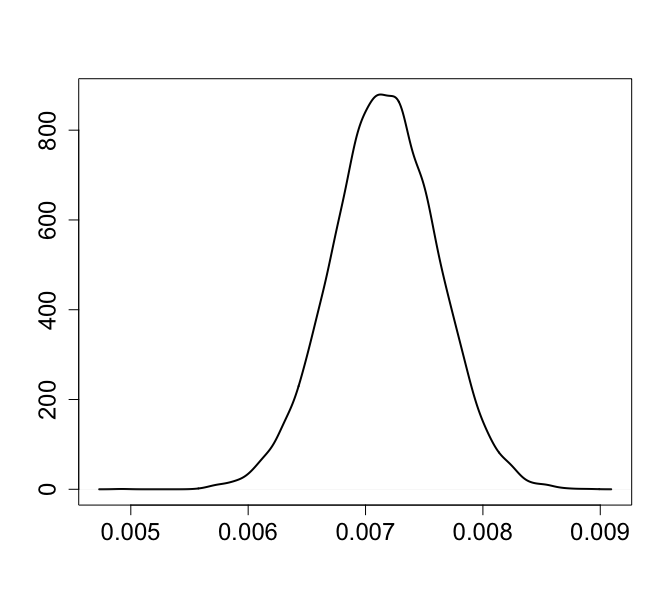}
}

\end{center}
\caption{Posterior distributions for the 7 parameters of the spatial ETAS model fitted to the South California earthquake catalog.}
\label{fig:results2}
\end{figure*}

As mentioned above, the 2010-2019  Southern California catalog contains a large magnitude 7.2 earthquake that occurred during July 2019. To evaluate forecast accuracy, we now fit the Bayesian ETAS model to the subset of the catalog which contains only this large earthquake, and those that occurred previously. This consisted of 1120 earthquakes and the resulting posterior distributions  for the $7$ parameters of the spatial ETAS model are shown in Figure \ref{fig:results2}.

Next, we produce an out-of-sample forecast for the total number of earthquakes that occurred after this large earthquake during the remainder of 2019. As a comparison, we produced a similar forecast using the standard (non-Bayesian) ETAS model where parameter uncertainty is ignored and the parameter vector $\theta$ is instead replaced by its maximum likelihood estimate. For both models, these forecasts were produced using the previously described simulation method to approximate the forecast distribution.

\begin{figure}[t]
\begin{center}
  \includegraphics[scale=0.35]{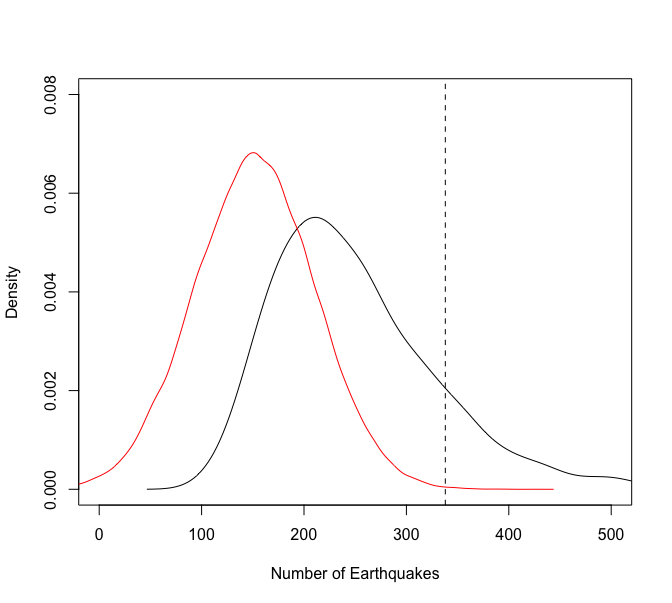}
  \end{center}
  \caption{Forecast distrbution for the number of magnitude $\geq 3.5$ earthquakes that occurred in 2019 after the large magnitude 7.2 mainshock in July. The black line is the forecast distribtion for the Bayesian ETAS model, while the red line is the forecast distribution for the standard ETAS model. The true number of earthquakes (338) is shown as a dotted vertical line}
  \label{fig:forecast2}
\end{figure}

Figure \ref{fig:forecast2} shows the resulting forecast distribution for the number of earthquakes using both the Bayesian and the non-Bayesian ETAS models. Note that the Bayesian version gives a substantially wider forecast distribution since it incorporates all uncertainty about the unknown parameter vector $\theta$. In contrast, the standard ETAS model does not incorporate this uncertainty since $\theta$ is replaced with the MLE. As such, the forecast distribution for the standard ETAS model is much narrower. It can be seen that the forecast distribution for the Bayesian ETAS model is consistent with the true number of earthquakes that occurred (338) while the forecast distribution for the standard ETAS model is not. This highlights that incorporating parameter uncertainty is important in order to avoid producing forecasts which are more confident than they have the right to be, which can result in flawed predictions. We note that some attempts to incorporate this uncertainty within a non-Bayesian framework have previously been discussed by \citep{fox_spatially_2016}.

Finally, we investigated the sensitivity of the Bayesian model to different choices of the prior distribution. Although our priors are intended to be non-informative, it is possible to construct such priors in different ways. For the Gamma prior on $\mu$ we tried specifications of Gamma($0.1,0.1$), Gamma($0.01,0.01)$, and Gamma($0.001,0.001)$, all of which correspond to varying degrees of non-informativity. For the other priors, we varied the upper bound of the Uniform distribution between $10$ and $1000$. As should be expected, none of these changes had any meaningful impact on the resulting forecast distribution, showing that the model is robust to reasonable changes to the priors.


\section*{Data and Resources}
Computer code implementing the Bayesian estimation framework introduced in this paper has been written in the language R, and is now available from CRAN  along with detailed instructions: https://cran.r-project.org/web/packages/bayesianETAS/index.html

The Southern California earthquake catalog which we analysed can be obtained from the SCEC data center at http://scedc.caltech.edu/research-tools/alt-2011-dd-hauksson-yang-shearer.html (Last accessed May 2020)

\bibliographystyle{apa}
\bibliography{jabref}

\end{document}